\documentstyle[prl,aps,preprint,amsmath,amsfonts,amssymb]{revtex}

\let\ssection=\section
\renewcommand{\section}{\setcounter{equation}{0}\ssection}

\def\a{\alpha}
\def\b{\beta}

\def\vta{\vartheta}

\input cyracc.def

\def\sqr#1#2{{\vcenter{\hrule height.#2pt\hbox{\vrule width.#2pt height.#1pt 
\kern#1pt \vrule width.#2pt}\hrule height.#2pt}}}

\def\hodge{{}^\star\!}

\def\Sigmakin{{}^{\rm k}\Sigma_\alpha}

\begin{document}
\title{On the energy-momentum current of the electromagnetic field in
  a pre-metric axiomatic approach\footnote{Dedicated to the memory of 
our friend and colleague {\it Ruggiero de Ritis} from Napoli.}. I}

\author{Friedrich W.\ Hehl\footnote{hehl@thp.uni-koeln.de} and 
Yuri N. Obukhov\footnote{yo@thp.uni-koeln.de, general@elnet.msk.ru}
\footnote{Department of Theoretical Physics, Moscow State University, 
117234 Moscow, Russia.}}
\address{Institute for Theoretical Physics, 
University of Cologne, 50923 K\"oln, Germany}


\maketitle 

\begin{abstract} 
{We complete a metric-free axiomatic framework for electrodynamics by
introducing the appropriate energy-momentum current $\Sigmakin$ of the
electromagnetic field. We start from the Lorentz force density and
motivate the form of $\Sigmakin$. Then we postulate it (fourth axiom)
and discuss its properties. In particular, it is found that
$\Sigmakin$ is traceless and invariant under an electric-magnetic
reciprocity transformation. By using the Maxwell-Lorentz spacetime
relation (fifth axiom), $\Sigmakin$ is also shown to be symmetric,
that is, it has 9 independent components.
}
\end{abstract}

\section{Introduction}

In an axiomatic approach to classical electrodynamics, one can start
with {electric charge conservation} $dJ=0$ as {\sl first} axiom, with
the existence of the {Lorentz force density}
$f_\alpha=(e_\alpha\rfloor F)\wedge J$ as {\sl second} axiom, and with
{magnetic flux conservation} $dF=0$ as {\sl third} axiom. Here $J$ is
the electric current density 3-form, $F=(E,B)$ the electromagnetic
field strength 2-form, and $e_\alpha$ the frame field, i.e., the
vector basis of the tangent space, with
$\alpha=\hat{0},\hat{1},\hat{2},\hat{3}$ as frame index. A fairly
detailed account, including the conventions and references to the
literature, can be found in \cite{Gentle}, see also \cite{HandO}; for
the mathematics involved, compare Frankel \cite{Ted}.

For the formulation of the axiomatics mentioned, one needs a
4-dimensional differentiable manifold as spacetime. It is assumed that
this manifold can be cut into 3-dimensional slices (folia) such that
successive slices can be numbered, at least locally, by means of a
monotone increasing parameter $\sigma$. The slices represent ordinary
3-space, the increasing parameter represents time. No linear
connection and no metric is used in this approach up to this stage.
Therefore we call it a {\em pre-metric} axiomatics (strictly, we
should also add the qualification ``pre-connection''). Accordingly,
our three axioms are valid on a sliceable, but otherwise arbitrary
4-dimensional manifold, see \cite{Puntigam}, be it described by means
of the Minkowskian geometry of special relativity (SR), the Riemannian
geometry of general relativity (GR), or the metric-affine geometry of
the metric-affine gauge theory of gravity (MAG, see \cite{PRs}).

The first axiom has the {\em inhomogeneous} Maxwell equation $dH=J$ as
a consequence, with $H=(\cal{D},\cal{H})$ as the electromagnetic
excitation 2-form. The third axiom represents the {\em homogeneous} Maxwell
equation $dF=0$. In order to complete the fundamental structure of
classical electrodynamics, one has to specify an axiom for the
distribution of the energy and the momentum and the corresponding
fluxes of the electromagnetic field. Only in this way we will be able
to find a formula for the electromagnetic radiation pressure, for
example. As it will turn out, this is also possible in a pre-metric
framework.

\section{Fourth axiom: localization of energy-momentum}

Let us consider the Lorentz force density $f_\alpha=(e_\alpha\rfloor
F)\wedge J$. If we want to derive the energy-momentum law for
electrodynamics, we have to try to express $f_\alpha$ as an {\em
  exact} form. Then energy-momentum is a kind of a generalized potential
for the Lorentz force density, namely $f_\alpha\sim d\,\Sigma_\alpha$. 
For that purpose we start from $f_\alpha$. We substitute $J=dH$ 
(inhomogeneous Maxwell equation) and subtract out a term with $H$ 
and $F$ exchanged and multiplied by a constant factor $a$:
\begin{equation}\label{inter1}
  f_\alpha = (e_\alpha\rfloor F)\wedge dH - a\, (e_\alpha\rfloor
  H)\wedge dF\,.
\end{equation}
Because of $dF=0$ (homogeneous Maxwell equation), the subtracted term
vanishes. The factor $a$ will be left open for the moment. Note that
we need a non-vanishing current $J\ne 0$ for our derivation to be
sensible.

We partially integrate both terms in (\ref{inter1}):
\begin{eqnarray}\label{inter2}
  f_\a&=& d[a\, F\wedge(e_\a\rfloor H) - H\wedge(e_\a\rfloor
  F){]}\nonumber\\ 
  &-&a\,F\wedge d(e_\a\rfloor H) + H\wedge d(e_\a\rfloor F)\,.
\end{eqnarray}
The first term has already the desired form. We recall the main
formula for the Lie derivative of an arbitrary form $\Phi$, namely $
{\cal L}_{e_\alpha} \Phi=d(e_\alpha\rfloor\Phi)+e_\alpha\rfloor
(d\Phi)\,.$ This allows us to transform the second line of
(\ref{inter2}):
\begin{eqnarray}\label{inter3}
  f_\a= d\bigl[a\,F\wedge(e_\a\rfloor H)&-&H\wedge(e_\a\rfloor
  F)\bigr]\nonumber\\ -a\,F\wedge ({\cal L}_{e_\a} H)&+&H\wedge ({\cal
    L}_{e_\a} F)\nonumber\\ +a\,F\wedge e_\a\rfloor (dH)&-&H\wedge
  e_\a\rfloor(d F)\,.
\end{eqnarray}
The last line can be rewritten as 
\begin{eqnarray}\label{inter4}
+a\,e_\a\rfloor[F\wedge dH]&-&a\,(e_\a\rfloor F)\wedge dH\nonumber\\
-e_\a\rfloor[H\wedge dF]&+&(e_\a\rfloor H)\wedge dF\,.
\end{eqnarray}
As 5-forms, the expressions in the square brackets vanish. Two terms
remain, and we find
\begin{eqnarray}\label{inter3a}
  f_\a= d\bigl[a\,F\wedge(e_\a\rfloor H)&-&H\wedge(e_\a\rfloor
  F)\bigr]\nonumber\\ -a\,F\wedge ({\cal L}_{e_\a} H)&+&H\wedge ({\cal
    L}_{e_\a} F)\nonumber\\ -a\,(e_\alpha\rfloor F)\wedge
    dH &+& (e_\alpha\rfloor H)\wedge
    dF \,.
\end{eqnarray}

Now we have to make up our mind about the choice of the factor $a$.
Because of $dF=0$, the third line adds up to $-af_\alpha$.
Accordingly,
\begin{eqnarray}\label{inter3b}
  (1+a)\,f_\a= d\bigl[a\,F\wedge(e_\a\rfloor H)&-&H\wedge(e_\a\rfloor
  F)\bigr]\nonumber\\ -a\,F\wedge ({\cal L}_{e_\a} H)&+&H\wedge ({\cal
    L}_{e_\a} F)\,.
\end{eqnarray}
With $a=-1$, the left hand side is zero and we find a mathematical
identity. A real conservation law is only obtained when, eventually,
the second line vanishes. In other words, here we need an a posteriori
argument, i.e., we have to take some information from experience. For
$a=0$, the second line does not vanish. However, for $a=1$, we can
hope that the first term in the second line compensates the second
term if somehow $H\sim F$. In fact, under ``ordinary circumstances'',
to be explored below, the two terms in the second line do compensate
each other for $a=1$. Therefore we postulate this choice and find
\begin{equation}
  f_\alpha = (e_\alpha\rfloor F)\wedge J =
  d\,\Sigmakin + X_\alpha\,.\label{fSX}
\end{equation}
Here the {\em kinematic energy-momentum} 3-form of the electromagnetic
field, a central result of this section, reads
\index{electromagnetic!energy-momentum!kinematic 3-form}
\begin{equation}
  ^ {\rm k} \Sigma_\alpha :={\frac 1 2}\left[F\wedge(e_\alpha\rfloor
    H) - H\wedge (e_\alpha\rfloor F)\right]\qquad\mbox{(fourth
    axiom)}\,,\label{simax}\end{equation} and the remaining force
density 4-form turns out to be
\begin{equation}  X_\alpha := -{\frac 1 2}\left(F\wedge{\cal
      L}_{e_\alpha}H-H\wedge{\cal L}_{e_\alpha}F \right)\,.\label{Xal}
\end{equation}
Incidentally, our formula (\ref{fSX}) supersedes the corresponding relation
of Post, see \cite{Post62} Eq. (4.42).

Our derivation of (\ref{fSX}) doesn't lead to a unique definition of
$\Sigmakin$. The addition of any closed 3-form would be possible,
\begin{equation}\label{nonunique}
  \Sigma_\a':=\Sigmakin +Y_\a\,,\qquad\text{with}\qquad
dY_\alpha=0\,,
\end{equation}
such that 
\begin{equation}\label{nonunique1}
  f_\a=d\,\Sigma_\a'+X_\a\,.
\end{equation}In particular, $Y_\alpha$ could be exact: 
$Y_\alpha=dZ_\alpha$. The 2-form $Z_\alpha$ has the same dimension as
$\Sigmakin$. It seems impossible to build up $Z_\alpha$ exclusively in
terms of the quantities $e_\alpha,\,H,\, F$ in an algebraic way.
Therefore, $Y_\a=0$ appears to be the most natural choice.  Thus, by the
{fourth axiom} we postulate that $\Sigmakin$ {\em in (\ref{simax})
  represents the energy-momentum current} that correctly localizes the
energy-momentum distribution of the electromagnetic field in
spacetime.

The current $\Sigmakin$ can also be rewritten by applying the
anti-Leibniz rule for $e_\a\rfloor$ either in the first or the second
term on the right hand side of (\ref{simax}). With the 4-form
\begin{equation}\label{lagr}\Lambda:=-\frac{1}{2}\,F\wedge H\,,
\end{equation}we find
\begin{eqnarray}\label{simax'}
  \Sigmakin &=&\!\quad e_\a\rfloor\Lambda
  +F\wedge(e_\alpha\rfloor H)\nonumber\\&=&-e_\a\rfloor
  \Lambda-H\wedge(e_\alpha\rfloor F)\,.
\end{eqnarray}

\section[Properties of energy-momentum, electric-magnetic 
reciprocity]{Properties of the energy-momentum
  current, electric-magnetic reciprocity}

\subsection{\bf $\Sigmakin$ is tracefree}

The energy-momentum current $\Sigmakin$ is a 3-form. We can blow it up
to a 4-form according to $\vartheta^\beta\wedge\Sigmakin$. Since it
still has 16 components, we haven't lost any information. If we recall
that for any $p$-form $\Phi$ we have $\vartheta^\a\wedge
(e_\a\rfloor\Phi)=p\Phi$, we immediately recognize from (\ref{simax})
that
\begin{equation}
\vartheta^\alpha\wedge\Sigmakin = 0\,,\label{zerotrace}
\end{equation}
which amounts to {\em one} equation. This property --- the vanishing
of the ``trace'' of $\Sigmakin$ --- is connected with the fact that
the electromagnetic field (the ``photon'') carries no mass and the
theory is thus invariant under dilations. Why we call that the trace
of the energy-momentum will become clear below, see (\ref{emtensor1}).

\subsection{\bf $\Sigmakin$ is electric-magnetic reciprocal}

Furthermore, we can observe another property of $\Sigmakin$. It is
remarkable how symmetric $H$ and $F$ enter (\ref{simax}). This was
achieved by our choice of $a=1$. The energy-momentum current is {\bf
  e}lectric-{\bf m}agnetic reciprocal, i.e., it remains invariant
under the {transformation}
\begin{equation}\label{duality1}
  H\rightarrow \zeta F\,,\quad F\rightarrow
  -\frac{1}{\zeta}\,H\,,\quad\Rightarrow\quad \Sigmakin\rightarrow
  \Sigmakin\,,
\end{equation}
with the twisted zero-form (pseudo-scalar function) $\zeta=\zeta(x)$
of dimension $[\zeta]=[H]/[F]=q^2/h$. Here $q$ denotes the dimension of
charge and $h$ that of action. 

It should be stressed that in spite of $\Sigmakin$ being e-m
reciprocal, Maxwell's equations are {\em not},
\begin{eqnarray}
dH=J\quad &\rightarrow&\quad  dF+F\wedge d\,\zeta/\zeta 
  =J/\zeta\,,\\dF=0\quad &\rightarrow&\quad dH-H\wedge
  d\,\zeta/\zeta = 0\,,
\end{eqnarray} 
not even for $d\,\zeta=0$, since we don't want to restrict ourselves 
to the free-field case with vanishing source $J=0$.

Eq.(\ref{duality1}) expresses a certain reciprocity between electric
and magnetic effects with regard to their respective contributions to
the energy-momentum current of the field. We call it electric-magnetic
reciprocity.\footnote{...following Toupin \cite{Toupin} even if he
  introduced this notion in a somewhat more restricted context.
  Maxwell spoke of the mutual embrace of electricity and magnetism,
  see Wise \cite{Wise}. In the case of a {\em prescribed metric},
  discussions of the corresponding Rainich ``duality rotation'' were
  given by Gaillard \& Zumino \cite{Gaillard} and by Mielke
  \cite{eggbuch}, amongst others.} That this naming is appropriate
can be seen from a (1+3)-decomposition. We recall, see \cite{Gentle},
the (1+3)-decompositions of $H$ and $F$:
\begin{equation}\label{sumhf}
H = -\,{\cal H}\wedge d\sigma + {\cal D}\,,\qquad F =\quad E\wedge
d\sigma + B\,.
\end{equation} 
We substitute them in (\ref{duality1}):
\begin{equation}\label{duality1a} 
  H\longrightarrow \!\zeta F \qquad
\begin{cases}{\cal H}\longrightarrow
    -\zeta E\,,\\ {\cal D}\longrightarrow \quad\zeta B\,,
\end{cases}
\end{equation}
\begin{equation}\label{duality1b} 
  F\longrightarrow -\frac{1}{\zeta}\,H
  \qquad\begin{cases}E\longrightarrow \quad\!\!\frac{1}{\zeta}\,{\cal
      H}\,,\\ B\longrightarrow -\frac{1}{\zeta}\,{\cal
      D}\,.\end{cases}\end{equation} 
Here it is clearly visible that a magnetic quantity is replaced by an
electric one {\em and} an electric quantity by a magnetic one:
$\text{\it electric}\longleftrightarrow \text{\it magnetic}$. In this
sense, we can speak of an electric-magnetic reciprocity in the
expression for the energy-momentum current $\Sigma_\alpha$.
Alternatively we can say that $\Sigma_\alpha$ 
fulfills e-m reciprocity, it is e-m reciprocal. 

Let us pause for a moment and wonder of how the notions ``electric''
and ``magnetic'' are attached to certain fields and whether there is a
conventional element involved. By making experiments with a cat's skin
and a rod of amber, we can ``liberate'' what we call {\em electric}
charges. In 3 dimensions, they are described by the charge density
$\rho$. Set in motion, they produce an electric current $j$.  The
electric charge is conserved (first axiom) and is linked, via the
Gauss law $\underline{d}\,{\cal D}=\rho$, to the {\em electric}
excitation $\cal D$.

Recurring to the Oersted experiment, it is clear that moving charges
$j$ induce magnetic effects, in accordance with the Oersted-Amp\`ere
law $\underline{d}\,{\cal H}-\dot{\cal D}= j$ --- also a consequence
of the first axiom. Hence we can unanimously attribute the term {\em
  magnetic} to the excitation $\cal H$. There is no room left for
doubt about that.

The second axiom links the electric charge density $\rho$ to the field
strength $E$ according to $(e_a\rfloor \rho)\wedge E$ and the electric
current $j$ to the field $B$ according to $(e_a\rfloor j)\wedge B$.
Consequently, also for the field strength $F$, there can be no other
way than to label $E$ as {\em electric} and $B$ as {\em magnetic}
field strength.

These arguments imply that the substitutions
$H\longrightarrow\zeta\,F$ as well as $F\longrightarrow -H/\zeta$ both
substitute an electric by a magnetic field and a magnetic by an
electric one, see (\ref{duality1a}) and (\ref{duality1b}). Because of
the minus sign (that is, because of $a=1$) that we found in
(\ref{duality1}) in analyzing the electromagnetic energy-momentum
current $\Sigma_\alpha$, we can{\em \/not} speak of an equivalence of
electric and magnetic fields, the expression reciprocity is much more
appropriate. Fundamentally, electricity and magnetism enter into
classical electrodynamics in an {\em a\/}symmetric way.

\medskip

  Let us try to explain the electric-magnetic reciprocity by means of
  a simple example. If we apply a $(1+3)$-decomposition to $\Sigmakin$,
  the electric energy density turns out to be $u_{\text{el}}=\frac{1}{2}
  \,E\wedge{\cal D}$.
  If one wants to try to guess the corresponding expression for the
  magnetic energy density $u_{\text{mag}}$, one substitutes for an
  electric a corresponding magnetic quantity. However, the electric
  field strength is a 1-form. One cannot substitute it by the magnetic
  field strength $B$ since that is a 2-form. Therefore one has to
  switch to the magnetic excitation according to $E\rightarrow
  \frac{1}{\zeta}\,{\cal H}$, with the 1-form ${\cal H}$. The function
  $\zeta$ is needed because of the different dimensions of $E$ and
  $\cal H$ and since $E$ is an untwisted and ${\cal H}$ a twisted
  form. Analogously, one substitutes ${\cal D}\rightarrow \zeta B$
  thereby finding $u_{\text{mg}}=\frac{1}{2}\,{\cal H}\wedge B$. This
  is the correct result, i.e., $u=\frac{1}{2}(E\wedge{\cal D} +
  B\wedge{\cal H})$, and we could be happy.

  Naively, one would then postulate the invariance of $u$ under the
  substitution $E\rightarrow \frac{1}{\zeta}\,{\cal H}\,,\,{\cal
    D}\rightarrow \zeta B\,,\,B\rightarrow \frac{1}{\zeta}\,{\cal
    D}\,,\,{\cal H}\rightarrow\zeta E$. But, as a look at (\ref{sumhf}) 
 will show, this cannot be implemented in a covariant way
  because of the minus sign in (\ref{sumhf}). One could reconsider the
  sign convention for ${\cal H}$ in (\ref{sumhf})$_1$. However, as a
  matter of fact, the relative sign between (\ref{sumhf})$_1$ and
  (\ref{sumhf})$_2$ is basically fixed by Lenz's rule (the induced
  electromotive force [measured in volt] is opposite in sign to the
  inducing field). Thus the minus sign in (\ref{sumhf}) is independent 
 of conventions.

  How are we going to save our rule of thumb for extracting the
  magnetic energy from the electric one? Well, if we turn to the
  substitutions (\ref{duality1a}) and (\ref{duality1b}), i.e., if we
  introduce two minus signs according to $E\rightarrow
  \frac{1}{\zeta}\,{\cal H}\,,\,{\cal D}\rightarrow \zeta
  B\,,\,B\rightarrow -\frac{1}{\zeta}\,{\cal D}\,,\,{\cal
    H}\rightarrow -\zeta E$, then $u$ still remains invariant and we
  recover the covariant rule (\ref{duality1}). In other words, the
  naive approach works {\em up to two minus signs}. Those we can
  supply by having insight into the covariant version of
  electrodynamics. Accordingly, the electric-magnetic reciprocity is
  the one that we knew all the time -- we just have to be careful with
  the sign.
\medskip

\subsection{$\Sigmakin$ expressed in terms of the complex 
  electromagnetic field}

We can understand the e-m reciprocity transformation as acting on the
column vector consisting of $H$ and $\zeta F$:
\begin{equation}\label{column}
\left(\begin{array}{c}H'\\ \zeta F'
  \end{array}\right)=
\left(\begin{array}{rr} 0\quad & 1\\-1\quad & 0
\end{array}\right) 
\left(\begin{array}{c}H\\ \zeta F
  \end{array}\right)\,.
\end{equation}
In order to compactify this formula, we introduce the {\em complex
  electromagnetic field}
\begin{equation}\label{complex}
  U:=H+i\,\zeta F\qquad{\rm and} \qquad U^*=H-i\,\zeta F\,,
\end{equation}with $^*$ denoting the conjugate complex.
Now the e-m reciprocity (\ref{column}) translates into
\begin{equation}\label{complex1}
U'=-i\,U\,,\qquad {U^{*}}'=i\,U^{*}\,.
\end{equation}
This corresponds, in the complex plane, where $U$ lives, to a rotation
by an angle of $-{\pi}/{2}$.

We can resolve (\ref{complex}) with respect to
excitation and field strength:
\begin{equation}\label{complex2}
H=\frac{1}{2}\,(U+U^*)\,,\qquad F=-\frac{i}{2\zeta}\,(U-U^*)\,.
\end{equation}
We differentiate (\ref{complex})$_1$. Then the Maxwell equation for
the complex field turns out to be
\begin{equation}\label{complexmax}dU+(U^*-U)\,\frac{d\zeta}{2\zeta} 
  =J\,.\end{equation} Clearly, if we choose a constant $\zeta$, i.e.,
$d\zeta=0$, the second term on the left hand side vanishes. The
asymmetry between electric and magnetic fields finds its expression in
the fact that the source term on the right hand side of
(\ref{complexmax}) is a {\em real} quantity.  

If we substitute (\ref{complex2}) into the energy-momentum current
(\ref{simax}), we find, after some algebra,
\begin{equation}\label{simaxcomplex}
  \Sigmakin=\frac{i}{4\zeta}\bigl[U^*\wedge(e_\alpha\rfloor U)-U
  \wedge(e_\alpha\rfloor U^*) \bigr]\,.
\end{equation}
Now, according to (\ref{complex1}), e-m reciprocity of the
energy-momentum current is manifest.

If we execute successively {\em two} e-m reciprocity transformations,
namely $U\longrightarrow U'\longrightarrow U''$, then, as can be seen
from (\ref{complex1}) or (\ref{duality1}), we find a reflection (a
rotation of $-\pi$), namely $U''=-U$, i.e.,
\begin{equation}\label{2duality}
  U\rightarrow -U\qquad\text{or}\qquad(H\rightarrow
  -H,\>F\rightarrow -F)\,.
\end{equation} Only {\em four}  e-m reciprocity transformations
lead back to the identity.  It should be stressed, however, that
already {\em one} e-m reciprocity transformation leaves $\Sigmakin$
invariant.

It is now straightforward to formally extend the e-m reciprocity
transformation (\ref{complex1}) to 
\begin{equation}\label{complex1a}
  U'=e^{+i\phi}\,U\,,\qquad {U^{*}}'=e^{-i\phi}\,U^{*}\,,
\end{equation}
with $\phi=\phi(x)$ as an arbitrary ``rotation'' angle. The
energy-momentum current $\Sigmakin$ is still invariant under this
extended transformation, but in later applications only the subcase of
$\phi=-\pi/2$, treated above, will be of interest.

\subsection{\bf Energy-momentum tensor density 
  ${}^{\rm k}{\cal T}_\alpha{}^\beta$}

Since $\Sigmakin$ is a 3-form, we can decompose it either conventionally 
or with respect to the basis 3-form $\epsilon_\beta:=e_\beta\rfloor\epsilon$,
with $\epsilon:= \vartheta^0 \wedge\vartheta^1 \wedge\vartheta^2
\wedge\vartheta^3$
:
\begin{equation}\label{emtensor}
  \Sigmakin=\frac{1}{3!}\,{}^{\rm k}\Sigma_{\lambda\mu\nu\alpha}\,
  \vartheta^\lambda\wedge\vartheta^\mu\wedge\vartheta^\nu = {}^{\rm
    k}{\cal T}_\alpha{}^\beta\,\epsilon_\beta\,.\end{equation} The 2nd
rank tensor density of weight 1, ${}^{\rm k}{\cal T}_\alpha{}^\beta$,
is the {\em Minkowski} energy tensor density. We can resolve this
equation with respect to ${}^{\rm k}{\cal T}_\alpha{}^\beta$ by exterior
multiplication with $\vartheta^\beta$. We recall
$\vartheta^\beta\wedge\epsilon_\gamma=\delta^\beta_\gamma\,\epsilon$
and find
\begin{equation}\label{emtensor1}
  {}^{\rm k}{\cal T}_\alpha{}^\beta\,\epsilon=\vartheta^\beta\wedge
  \Sigmakin\,.
\end{equation}
Thereby we recognize that $\vartheta^\alpha\wedge \Sigmakin=0$, see
(\ref{zerotrace}), is equivalent to the vanishing of the trace of the
energy-momentum tensor density ${}^{\rm k}{\cal T}_\alpha{}^\alpha =0$. 
Thus ${}^{\rm k}{\cal T}_\alpha{}^\beta$ as well as $\Sigmakin$ have 
15 independent components at this stage. Both quantities are equivalent.

If we substitute (\ref{simax}) into (\ref{emtensor1}), then we can
express the energy-momentum tensor density in the components of $H$
and $F$ as follows\footnote{Our conventions with respect to the
totally antisymmetric Levi-Civita symbol are as follows: We have 
$\epsilon^{\hat{0}\hat{1}\hat{2}\hat{3}} = +1$, i.e., 
$\epsilon^{\alpha\beta\gamma\delta}$ is a tensor density of weight 
$+1$. Its reciprocal $\hat{\epsilon}_{\mu\nu\rho\sigma}$, a tensor 
density of weight $-1$, is defined according to $\epsilon^{\alpha
\beta\gamma\delta}\,\hat{\epsilon}_{\mu\nu\rho\sigma}=\delta^{\alpha
\beta\gamma\delta}_{\mu\nu\rho\sigma}$, with $\delta$
as generalized Kronecker symbol.}: 
\begin{equation}\label{emtensor2}
  {}^{\rm k}{\cal T}_\alpha{}^\beta=\frac{1}{4}\,\epsilon^{\beta
    \mu\rho\sigma}\left(H_{\alpha\mu}F_{\rho\sigma}-F_{\alpha\mu}H_{\rho
      \sigma} \right)\,.
\end{equation}

\subsection{Preview: Covariant conservation law and vanishing 
force density $\widehat{X}_\alpha$}

The Lorentz force density $f_\alpha$ in (\ref{fSX}) and the
energy-momen\-tum current $\Sigmakin$ in (\ref{simax}) are covariant
with respect to frame and coordinate transformations.  Nevertheless,
each of the two terms on the right hand side of (\ref{fSX}), namely
$d\,\Sigmakin$ or $X_\a$, are {\em not} covariant by themselves. What
can we do?

For the first three axioms of electrodynamics, the spacetime arena is
only required to be a (1+3)-decomposable 4-dimensional manifold. We
cannot be as economical as this in general. Ordinarily a {\em linear
  connection} $\Gamma_\alpha{}^\beta$ on that manifold is needed.  The
linear connection will be the guiding field that transports a vector,
e.g., from one point of spacetime to a neighboring one. Then the 
covariant exterior differential is be defined $D=d + \Gamma_{\alpha} 
{}^{\beta}\rho(L_\b{}^\a)$, see \cite{PRs}. With the help of this operator, 
a generally covariant expression $D\,\Sigmakin$ can be constructed. 
Then (\ref{fSX}) can be rewritten as
\begin{equation}f_\alpha = D\,\Sigmakin + 
  \widehat{X}_\alpha\,,\label{fSXgam}
\end{equation}
with the new supplementary force density
\begin{equation}
\widehat{X}_\alpha = {\frac 1 2}\left(H\wedge{\hbox{\L}}_{e_\alpha}F -
F\wedge{\hbox{\L}}_{e_\alpha}H\right)\,,\label{Xalgam}
\end{equation}
which contains the gauge covariant Lie derivative 
\begin{equation}\label{magic}
\hbox{\L}_\xi := D\,\xi\rfloor + \xi\rfloor D\,.
\end{equation}
Note that the energy-momentum current $\Sigmakin$ remains the same,
only the force density $X_\alpha$ is replaced by $\widehat{X}_\alpha$.  
Now we will be able to show that the fourth axiom is exactly what is 
needed for an appropriate and consistent derivation of the conservation 
law for energy-momentum.

It is remarkable, in (\ref{fSXgam}) [or in (\ref{fSX})] the
energy-momen\-tum current can be defined even if (\ref{fSXgam}), as
long as $\widehat{X}_\alpha\ne 0$, doesn't represent a conservation
law.  However, in order to find a genuine conservation law, we have to
require that $\widehat{X}_\alpha$ vanishes. This cannot be achieved
without some knowledge on the relation between excitation and field
strength. Only this {\em electromagnetic spacetime relation} between
$H$ and $F$, which we shall postulate as fifth axiom in (\ref{axiom5}), 
makes electrodynamics a complete theory. At first, however, we don't 
want to commit ourselves to this axiom. We rather would like to exploit 
the arbitrary linear connection $\Gamma_\alpha{}^\beta$, introduced 
above, as far as possible. As auxiliary quantities,
attached to $\Gamma_\alpha{}^\beta$, we need the {\em torsion} 2-form
$T^\alpha:=D\vartheta^\alpha=d\vartheta^\alpha+\Gamma_\beta{}^\alpha
\wedge\vartheta^\beta=T_{\mu\nu}{}^\alpha\,\vartheta^ \mu \wedge
\vartheta^\nu/2$ and the {\em transposed connection} 1-form
${\buildrel\frown\over{\Gamma}}_{\alpha}{}^{\beta}:=
\Gamma_{\alpha}{}^{\beta} +e_{\alpha}\rfloor T^{\beta}$.

Let us now consider the extra (supplementary, or offending) force
density $\widehat{X}_\alpha$. 
What we need is the gauge covariant Lie derivative of
an arbitrary 2-form
$\Psi=\Psi_{\mu\nu}\,\vartheta^\mu\wedge\vartheta^\nu/2$ in terms of
its components. Using (\ref{magic}), some algebra yields
\begin{equation}
  \hbox{\L}_{e_\alpha}\Psi ={\frac 1 2}\,
  \left({\buildrel\frown\over{D}}_\alpha
  \Psi_{\mu\nu}\right)\,\vartheta^\mu\wedge\vartheta^\nu \,,
\end{equation}
where ${\buildrel\frown\over{D}}_\alpha :=
e_\alpha\rfloor{\buildrel\frown\over{D}}$, with
${\buildrel\frown\over{D}}$ as the exterior covariant differential
with respect to the transposed connection. Thus,
 \begin{equation}
  \widehat{X}_\alpha = {\frac 1 8}
 \left(H_{\rho\sigma}{\buildrel\frown\over{D}}_\alpha F_{\mu\nu} -
      F_{\rho\sigma}{\buildrel\frown\over{D}}_\alpha H_{\mu\nu}\,
    \right) \vartheta^\rho \wedge\vartheta^\sigma
    \wedge\vartheta^\mu\wedge\vartheta^\nu\,,
\end{equation}
or, since $\vartheta^\rho\wedge\vartheta^\sigma\wedge\vartheta^\mu\wedge
\vartheta^\nu = \epsilon^{\rho\sigma\mu\nu}\,\epsilon\,$,\, we find
 \begin{equation}\label{extra}
   \widehat{X}_\alpha = {\frac \epsilon 8}\,\epsilon^{\rho\sigma\mu\nu}
   \left(H_{\rho\sigma}{\buildrel\frown\over{D}}_\alpha F_{\mu\nu} -
     F_{\rho\sigma}{\buildrel\frown\over{D}}_\alpha H_{\mu\nu}\,
   \right)\,.\end{equation}
This is as far as we can go with an arbitrary linear connection.

\section{Linear spacetime relation, Abelian axion}\label{5-MLA}

In order to make electrodynamics a complete theory, eventually more
detailed properties of spacetime have to come into play. We have to
find a relationship between excitation $H$ and field strength $F$. We
shall call it the {\em electromagnetic spacetime relation}.\footnote{
Truesdell \& Toupin \cite{TT}, Toupin \cite{Toupin}, and Kovetz 
\cite{Attay} call it ``aether relations''.} Following Toupin 
\cite{Toupin}, Sch\"onberg \cite{Schoen71}, Jadczyk \cite{Jadczyk}, 
and the authors with Rubilar and Fukui \cite{OH,OFR}, see also Gross 
\& Rubilar \cite{GR}, a {\em linear} relation is assumed between $H$ 
and $F$. In coordinate (natural) components, it reads
\begin{equation}\label{chiHF}
H_{ij} =  \frac{1}{2}\,\kappa_{ij}{}^{kl}\,F_{kl}\,.
\end{equation}
Compare also Tamm \cite{Tamm} and Post \cite{Postchi} in this context.
This linear spacetime relation postulates the existence of $6\times
6$ pseudo-scalar functions $\kappa$ with
\begin{equation}\label{oper6}
  \kappa_{ij}{}^{kl}=-\, \kappa_{ji}{}^{kl}=-\,
  \kappa_{ij}{}^{lk}\,.\end{equation} 
We can alternatively introduce
\begin{equation}\label{raise}
\chi^{ijkl}=
  \frac{1}{2}\,\epsilon^{ijmn}\,\kappa_{mn}{}^{kl}\,.\end{equation}
Its reciprocal reads
\begin{equation}\label{lower}
  \kappa_{ij}{}^{kl}=\frac{1}{2}\,\hat{\epsilon}_{ijmn}\,\chi^{mnkl} \,.
\end{equation}
The 36 functions $\kappa_{ij}{}^{kl}(\sigma,x)$ as well as the
$\chi^{ijkl}(\sigma,x)$ depend on time $\sigma$ and on space $x$ in
general. Because of the corresponding properties of the Levi-Civita
symbol, the $\chi^{ijkl}$ represent an untwisted {\it tensor density
  of weight $+1$}.  As is clear from (\ref{chiHF}) and (\ref{lower}),
the $\kappa_{ij}{}^{kl}$ and the $\chi^{ijkl}$ both carry the
dimension $[\kappa]=[\chi]=q^2/h$.

With the linear law we can set up a Lagrange 4-form $V$.
Because of $H=-\,\partial V/\partial F$, the Lagrangian must be
quadratic in $F$.  Thus we find
\begin{eqnarray}
  V&=&-\,\frac{1}{2}\,H\wedge F=-\,\frac{1}{8}\,H_{ij}F_{pq}\,dx^i\wedge
  dx^j\wedge dx^p\wedge dx^q\nonumber\\ 
  &=&-\,\frac{1}{32}\,\chi^{mnkl}\,\hat{\epsilon}_{ijmn}\,F_{kl}
  F_{pq}\,dx^i\wedge dx^j\wedge dx^p\wedge dx^q\,.
\end{eqnarray}
We rewrite the exterior products with the Levi-Civita symbol. Then we have
\begin{equation}
  V=-\,\frac{1}{8}\,\chi^{ijkl}\,
   F_{ij}F_{kl}\,dx^0\wedge dx^1\wedge dx^2\wedge dx^3\,.
\end{equation}
The components of the field strength $F$ enter in a {\em symmetric}
way. Therefore, without loss of generality, we can impose the
symmetry condition
\begin{equation}
  \chi^{ijkl} = \chi^{klij}
  \label{sym}
\end{equation}
reducing them to 21 independent functions at this stage.

Right now (still without a metric), we can split off the totally 
antisymmetric part of $\chi^{ijkl}$ 
according to
\begin{equation}\label{split}
\chi^{ijkl}= {\widetilde{\chi}}{\,}^{\!ijkl}+\alpha\,\epsilon^{ijkl}\,, 
\qquad{\rm  with}\qquad {\widetilde{\chi}}{\,}^{\![ijkl]}=0\,.
\end{equation}
Here $\alpha=\alpha(\sigma,x)$ is a pseudo-scalar
function with the dimension $[\alpha]=q^2/h$.
Thus the {\em linearity ansatz} eventually reads
\begin{equation}\label{linear}
H_{ij}= \frac{1}{4}\,\hat{\epsilon}_{ijmn}\,\chi^{mnkl}
\,F_{kl}= \frac{1}{4}\hat{\epsilon}_{ijmn}
\,{\widetilde{\chi}}{\,}^{\!mnkl}\,F_{kl}+\alpha\,F_{ij}\,,
\end{equation}
with 
\begin{equation}\label{chisymm}
{\widetilde{\chi}}{\,}^{mnkl}=-{\widetilde{\chi}}{\,}^{nmkl}
=-{\widetilde{\chi}}{\,}^{mnlk}={\widetilde{\chi}}{\,}^{klmn}
\quad{\rm and}\quad {\widetilde{\chi}}{\,}^{[mnkl]}=0\,.
\end{equation}
Besides the {\it Abelian axion} field $\alpha$, we have 20 independent
functions. Accordingly, the tensor ${\widetilde{\chi}}{\,}^{nmkl}$
has the same algebraic symmetries and the same number of independent
components as a curvature tensor in a 4-dimensional Riemannian
spacetime. It is remarkable that the pseudo-scalar axion field 
$\alpha$, see \cite{Ni} and references therein, enters here as a
quantity that does not interfere at all with the first four axioms of 
electrodynamics. Already at the pre-metric level, such a field emerges 
as a not unnatural companion of the electromagnetic field. However, to
make it a real independent field, kinetic terms $\sim d\alpha$ would
have to be added to the Lagrange 4-form $V$. 

Alternatively, the linear relation (\ref{linear}) can be reformulated as 
\begin{equation}\label{linear1}
H_{ij}= \frac{1}{2}\,\widetilde{\#}_{ij}{}^{kl}\,F_{kl} + \alpha\,F_{ij}\,,
\end{equation}
where $\widetilde{\#}$ (speak ``sharp tilde'') is a certain 
{\em duality} operator which acts linearly in the space of 
two-forms,\index{duality operator}
\begin{equation}
{}^{\widetilde{\#}}:\,\Lambda^2X\longrightarrow\Lambda^2X. 
\end{equation}
In particular, the action on the basis of 2-forms reads:
\begin{equation}\label{sharp2}
{}^{\widetilde{\#}}\left(dx^i\wedge dx^j\right)=\frac{1}{2}
\,{\widetilde{\#}}_{kl}{}^{ij}\left(dx^k\wedge dx^l\right) =
\frac{1}{4}\,\epsilon_{klmn}\,{\widetilde{\chi}}{\,}^{mnij}
\left(dx^k\wedge dx^l\right).
\end{equation}
We display the duality operator as a superscript ${}^{\widetilde{\#}}$. 
However, in its components ${\widetilde{\#}}_{ij}{}^{kl}$, the sharp 
sign is treated like an ordinary letter.

Now the linear material law (\ref{linear}) can be written as
\begin{equation}\label{lin2}
H=\left({}^{\widetilde{\#}} +\alpha\right)F. 
\end{equation} 

\section{Electric-magnetic reciprocity of the spacetime relation}\label{emsr}

If we exploit the linear relationship between excitation and field
strength and substitute it in the Maxwell equations, we end up \cite{OFR}
with the propagation of the electromagnetic disturbances in vacuum 
spacetime along {\it quartic wave surfaces}, for related work see 
\cite{Ditt,Lor,Leon}. Accordingly, spacetime would be {\em 
birefringent\,}, for instance, a property which seems to be contrary 
to experiment, at least at in our part of the universe. If, say, at 
a very early time in the development of the universe the vacuum were 
birefringent, then we had already a consistent theory for such an 
effect, namely the one we formulated in the last section.

If we want to forbid birefingence for vacuum spacetime, then we can
try to impose a {\it constraint} on the linear relationship (\ref{lin2}). 
The obvious choice is to require electric-magnetic reciprocity for
(\ref{lin2}). We have discovered e-m reciprocity as a property of
the energy-momentum current $\Sigmakin$ of the electromagnetic field.
Why should't we apply it to (\ref{lin2})? 

The e-m reciprocity transformation
\begin{equation}\label{duality2}
  H\rightarrow \zeta F\,,\qquad F\rightarrow
  -\frac{1}{\zeta}\,H\,,
\end{equation}
can alternatively be written as
\begin{equation}\label{duality2a}
  \left(\begin{array}{c}H\\F\end{array}\right)\rightarrow
    \left(\begin{array}{cc}0{\ }&{\ }\zeta\\-\frac{1}{\zeta}{\ } &{\ }0
      \end{array}\right)\left(\begin{array}{c}H\\F\end{array}\right)= 
        \left(\begin{array}{c}\zeta
            F\\-\frac{1}{\zeta}\,H\end{array}\right)\,.
\end{equation}
If 
\begin{equation}\label{duality2b}
W:= \left(\begin{array}{cc}0{\ }&{\ } \zeta\\-\frac{1}{\zeta}{\ } &{\ }0
      \end{array}\right)\,,\quad\text{then}\quad W^{-1}=
 \left(\begin{array}{cc}0{\ }&{\ }-\zeta\\ \frac{1}{\zeta}{\ } &{\ } 0
      \end{array}\right)\,.
\end{equation}
Let us perform an e-m reciprocity transformation in (\ref{lin2}). By
definition, the reciprocity transformation {\em commutes} with the 
duality operator ${}^{\widetilde{\#}}$. Then we find
\begin{equation}\label{duality3}
  \zeta F= \left({}^{\widetilde{\#}} +\alpha
  \right)\left(-\frac{H}{\zeta}\right)\quad\text{or}\quad 
  {}^{\widetilde{\#}}H=-\zeta^2\,F - \alpha\,H\,.
\end{equation}
On the other hand, we can also apply the duality operator 
${}^{\widetilde{\#}}$ to (\ref{lin2}). Because ${}^{\widetilde{\#}}$ 
commutes with 0-forms, we get
\begin{equation}\label{lin3} 
{}^{\widetilde{\#}}H=\left({}^{\widetilde{\#}\widetilde{\#}} 
+\alpha\,{}^{\widetilde{\#}}\right)F\,.
\end{equation} 

If we postulate e-m reciprocity of the linear law (\ref{lin2}),
then, as a comparison of (\ref{lin3}) and (\ref{duality3}) shows, 
we have to assume additionally
\begin{equation}\label{close1}
{}^{\widetilde{\#}}{}^{\widetilde{\#}}=-\zeta^2\,{\mathbb I},
\qquad\qquad \alpha=0\,.
\end{equation}
We call ${}^{\widetilde{\#}}{}^{\widetilde{\#}}=-\zeta^2{\mathbb I}$ the 
{\it closure relation} since applying the operator ${}^{\widetilde{\#}}$ 
twice, we come back to the identity operator ${\mathbb I}$ (= 
$\delta^{ij}_{kl}$, in components), up to a negative function. In this 
sense, the operation closes.\index{duality operator!closure condition}
At the same time, (\ref{close1}) tells us that the spacetime relation is 
{\it not} e-m reciprocal for an arbitrary transformation function $\zeta$ 
(like the energy-momentum current is). The duality operator 
${}^{\widetilde{\#}}$ is based on the measurable components 
$\widetilde{\chi}{\,}^{ijkl}$. If applied twice, as in (\ref{close1}),
there must not emerge an arbitrary function. In other words, we can
solve (\ref{close1})$_1$, by taking its trace, to get
\begin{equation}
\zeta^2 = -\,{\frac 1 {6}}\,{\rm Tr}\left({}^{\widetilde{\#}}
{}^{\widetilde{\#}}\right) =  -\,{\frac 1 {24}}\,\widetilde{\#}_{kl}{}^{ij}
\,\widetilde{\#}_{ij}{}^{kl} =: \lambda^2\,.
\end{equation}

It is natural to factorize the dimensionfull function $\lambda$ in the
constitutive matrix (\ref{split})-(\ref{linear1}) which yields a 
representation
\begin{equation}
\widetilde{\chi}{\,}^{ijkl} = \lambda\stackrel{\rm o}{\chi}{\!}^{ijkl}
\end{equation}
so that $\stackrel{\rm o}{\chi}{\!}^{ijkl}$ is now the
{\it dimensionless} tensor with the same symmetries as in (\ref{chisymm}).
This effectively redefines the duality operator
\begin{equation}\label{close1a'}
^{\widetilde{\#}}=:\lambda\,{}^{\#} \,,
\end{equation}
and for ${}^{\#}$ the closure reads 
\begin{equation}\label{close1a}
  ^{\#\#}=-\,{\mathbb I}.
\end{equation}
As we will see, the minus sign is very decisive: It will eventually
yield the Lorentzian signature of the metric of spacetime.

Let us collect our results. The e-m reciprocity of the linear
ansatz leads to the relations
\begin{equation}\label{collectHF}
H=\lambda\,^{\#} F\qquad\text{and}\qquad ^{\#}H=-\lambda\,F\,.
\end{equation}

We can also consider the duality operator ${}^\#$ from another point
of view. It is our desire to eventually describe empty spacetime
with such a linear ansatz. Therefore we have to reduce the number of 
independent functions $\stackrel{\rm o}{\chi}{\!}^{ijkl}$ somehow. 
The only constants with even parity are the Kronecker deltas $\delta_i^j$.
Obviously the $\delta^j_i$'s are of no help here in specifying the
$\stackrel{\rm o}{\chi}{\!}^{ijkl}$'s, since they carry also lower 
indices which cannot be absorbed in a non-trivial way in order to 
bring about the 4 upper indices of $\stackrel{\rm o}{\chi}{\!}^{ijkl}$. 
Recognizing that in the framework of electrodynamics in matter a similar 
linear ansatz can describe anisotropic media, we need a condition in 
order to exclude so-called non-reciprocal effects. A ``square'' of 
$\stackrel{\rm o}{\chi}$ will do the job,
\begin{equation}\label{close3}
\frac{1}{8}\,\stackrel{\rm o}{\chi}{}^{ijkl}\,\stackrel{\rm o}
{\chi}{}^{mnpq}\,\hat{\epsilon}_{klmn}\,\hat{\epsilon}_{pqrs}
= \mp\,\delta^{ij}_{rs}\,,
\end{equation}
with $\delta^{ij}_{rs}$ as a generalized Kronecker delta. This equation 
represents, together the linearity ansatz, our fifth axiom for spacetime. 
 Alternatively, it can also be written as 
 \begin{equation}\label{close4}
{\frac 1 2}\,\#_{ij}{}^{kl}\, \#_{kl}{}^{mn}=\mp\,\delta_{i\,j}^{mn}\,.
 \end{equation}
 Apparently, the equations (\ref{close1}) or (\ref{close3}) and
 (\ref{close4}), being consequences of e-m reciprocity, are just 
alternative formulations of the closure property.

\section{Fifth axiom, the construction of the metric}

The linear spacetime relation $H = (\lambda\,^{\#} + \alpha)F$, see 
(\ref{lin2}) and (\ref{close1a'}), together with the constraints 
$^{\#\#} = -1$ and $\alpha =0$, see (\ref{close1}), constitute our 
fifth axiom of electrodynamics. Since the dimensions of $H$ and $F$ are 
fixed, the unknown scalar function $\lambda$ is to be determined by  
experiment (like all of the components of the electromagnetic spacetime
matrix). Thus our axiom reads
\begin{equation}
H=\lambda\,^{\#} F,\qquad\text{with}\qquad ^{\#\#} = -1. \label{axiom5}
\end{equation}

The closure relation (\ref{axiom5})$_2$ affects the quartic wave 
surface mentioned at the beginning of Sec.~\ref{emsr}. The wave surface
reduces to the usual light cone surface. Thus, eventually, the
closure relation yields the unique lightcone structure for the 
propagation of electromagnetic waves. As a result, up to a conformal 
factor, the {\it spacetime metric} $g$ with the correct 
Lorentzian signature is constructed by means of the fifth axiom.
Our duality operator becomes the Hodge star operator ${}^\star$ with
respect to the metric $g$, see \cite{OH,OFR,GR}, i.e., the fifth axiom
becomes 
\begin{equation}\label{constvac}
  H=\lambda\,\hodge F\,.
\end{equation}\index{constitutive law!linear!local}
For $\lambda = const$, we will call (\ref{constvac}) {\it the 
Maxwell-Lorentz spacetime relation}. Only a constant $\lambda$ will
provide for the vanishing of the extra force density (\ref{extra}).

The numerical value of the constant factor in (\ref{constvac}) is fixed by
experiment:  
\begin{equation}\label{lambda}
\lambda = \sqrt{\frac{\varepsilon_0} {\mu_0}}=
{\frac {e^2} {4\pi\alpha_{\rm f}\hbar}}= 2.6544187283\,{\frac 1 {k\,\Omega}}.
\end{equation} 
Here $e$ is the charge of the electron and $\alpha_{\rm f}=1/137.036$ 
the fine structure constant. The inverse $1/\lambda$ is called the 
characteristic impedance (or wave resistance) of the vacuum. This is a 
fundamental constant which describes the basic electromagnetic {\it property 
of spacetime} if considered as a special type of medium (sometimes called 
{\it vacuum}, or {\it aether}, in the old terminology). In this sense, 
one can understand (\ref{constvac}) as the constitutive relations for the 
spacetime itself. The Maxwell-Lorentz spacetime relation (\ref{constvac}) 
is {\it universal}. It is equally valid in Minkowski, Riemannian, and 
post-Riemannian spacetimes. 
The {\it electric constant} $\varepsilon_0$ and the {\it magnetic 
constant} $\mu_0$ (also called vacuum permittivity and vacuum permeability, 
respectively) determine the universal constant of nature
\begin{equation}
c = {\frac 1{\sqrt{\varepsilon_0\mu_0}}}\label{cem}
\end{equation}
that describes the velocity of light in vacuum. 

If (\ref{constvac}) is substituted into the Maxwell equations, we find 
the Maxwell-Lorentz  equations 
\begin{equation}\label{MaxLor}
d\,{}^\star\! F = J/\lambda\,,\qquad dF=0\end{equation}
of standard electrodynamics.

\section{Symmetry of the energy-momentum current of 
the electromagnetic field}\label{13MAXmom}

If the spacetime metric $g$ is given, then there exists a unique
torsion-free and metric-compatible Levi-Civita connection 
$\widetilde{\Gamma}_\alpha{}^\beta$. Consider the conservation law
(\ref{fSXgam}). In a Riemannian space, the covariant Lie derivative
$\widetilde{\hbox{\L}}_\xi = \widetilde{D}\,\xi\rfloor + \xi\rfloor
\widetilde{D}$ commutes with the Hodge operator, $\widetilde
{\hbox{\L}}_\xi{}^\star\,={}^\star\widetilde{\hbox{\L}}_\xi$. Thus
(\ref{Xalgam}) straightforwardly yields
\begin{equation}
  \widehat{X}_\alpha ={\frac \lambda 2}\left({}^\star F\wedge
    \widetilde{\hbox{\L}}_{e_\alpha}F -
    F\wedge\widetilde{\hbox{\L}}_{e_\alpha}{}^\star
    F\right)=0\,.\label{Xalriem}
\end{equation}

Therefore in {\it general relativity} (GR), with the Maxwell-Lorentz 
spacetime relation, (\ref{fSXgam}) simply reduces to
\begin{equation}
  \widetilde{D}\,\Sigmakin =(e_\alpha\rfloor F)\wedge J\,.
\end{equation}
The energy-momentum current (\ref{simax}) now reads
\begin{equation}
\Sigmakin =\frac{\lambda}{2}\bigl[F\wedge(e_\alpha\rfloor{}^\star F) 
- (e_\alpha\rfloor F)\wedge{}^\star F\bigr]\,.\label{maxmomergy}
\end{equation}
In the absence of sources, $J=0$, we find the energy-momentum law
\begin{equation}
\widetilde{D}\,\Sigmakin =0\,.
\end{equation}\index{electromagnetic!energy-momentum!conservation law}
In the {\it flat} Minkowski spacetime of SR, we can {\it globally} choose the
coordinates in such a way that $\widetilde{\Gamma}_\alpha{}^\beta =0$.
Thus $\widetilde{D}\stackrel{*}{=}d$ and $d\,\Sigmakin =0$.

As we already know from (\ref{zerotrace}), the current (\ref{maxmomergy}) 
is {\it traceless} $\vartheta^\alpha\wedge\Sigmakin =0$. Moreover, we 
now can use the metric and prove also its symmetry. We multiply 
(\ref{maxmomergy}) by $\vartheta_\beta = g_{\beta\gamma}\,\vartheta^\gamma$
and antisymmetrize:
\begin{eqnarray}
{\frac 4 \lambda}\,\vartheta_{[\beta}\wedge{}^{\rm k}\Sigma_{\alpha]} 
&=&\quad \vartheta_\beta\wedge F\wedge (e_\alpha\rfloor{}^\star F) - 
\vartheta_\beta\wedge (e_\alpha\rfloor F)\wedge {}^\star F \nonumber\\
&& -\,\vartheta_\alpha\wedge F\wedge (e_\beta\rfloor{}^\star F) +
\vartheta_\alpha\wedge (e_\beta\rfloor F)\wedge {}^\star F.\label{momergy1}
\end{eqnarray}
Because of the identities $e_\alpha\rfloor{}^\star\Phi = {}^\star (\Phi
\wedge\vartheta_\alpha)$ and ${}^\star\Phi\wedge\Psi = {}^\star\Psi\wedge
\Phi$ (for all $p$-forms $\Psi$ and $\Phi$), the first term on the 
right-hand side can be rewritten, 
\begin{eqnarray}
\vartheta_\beta\wedge F\wedge (e_\alpha\rfloor{}^\star F) &=& F\wedge
\vartheta_\alpha\wedge{}^\star\left(\vartheta_\beta\wedge F\right)
\nonumber\\ 
&=& F\wedge\vartheta_\alpha\wedge (e_\beta\rfloor {}^\star F),\label{momergy2}
\end{eqnarray}
i.e., it is compensated by the third term. We apply the analogous technique
to the second term. Because ${}^{\star\star}F = - F$, we have 
\begin{eqnarray}
\vartheta_\beta\wedge {}^\star\left({}^\star F\wedge\vartheta_\alpha\right)
\wedge {}^\star F &=& {}^\star\left({}^\star F\wedge\vartheta_\alpha\right)
\wedge\vartheta_\beta\wedge{}^\star F \nonumber\\
&=& - {}^\star\left(\vartheta_\beta\wedge {}^\star F\right)\wedge 
{}^\star F\wedge\vartheta_\alpha \nonumber\\ &=& 
- \vartheta_\alpha\wedge (e_\beta\rfloor F)\wedge {}^\star F.\label{momergy3}
\end{eqnarray}
In other words, the second term is compensated by the fourth one and we
finally have
\begin{equation}
\vartheta_{[\beta}\wedge{}^{\rm k}\Sigma_{\alpha]} = 0.\label{momergy4}
\end{equation}

Alternatively, we can work with the energy-momentum tensor. 
We decompose the 3-form $\Sigmakin$ with respect to the 
$\eta$-basis. This is now possible since a metric is available.
Because of $\vta^\a\wedge\eta_\gamma=\delta^\a_\gamma\,\eta$,
with the volume 4-form $\eta = \sqrt{-\det g_{\mu\nu}}\,\vartheta^{\hat{0}}
\wedge\vartheta^{\hat{1}}\wedge\vartheta^{\hat{2}}\wedge\vartheta^{\hat{3}}$, 
we find
\begin{equation}\label{EMTensor}
  \Sigmakin =: {}^{\rm k}T_\a{}^\b\,\eta_\b\,\qquad \text{or}\qquad
  {}^{\rm k}T_{\alpha\beta}={}^\star\left(\vta_\b\wedge \Sigmakin\right)\,,
\end{equation} 
compare this with (\ref{emtensor})-(\ref{emtensor1}). We have
\begin{equation}
{}^{\rm k}T_{\a\b}={}^{\rm k}{\cal T}_{\a\b}/\sqrt{-g}.
\end{equation}
Its tracelessness ${}^{\rm k}T_\gamma{}^\gamma =0$ has already been 
established, see (\ref{emtensor1}), its symmetry
\begin{equation}
  {}^{\rm k}T_{[\a\b]}=0\label{maxsym}
\end{equation}
can be either read off from (\ref{EMTensor}) and (\ref{momergy4}) 
or directly from (\ref{emtensor2}) with ${\cal H}^{ij} =
{\frac 12}\,\epsilon^{ijkl}\,H_{kl}\sim F^{ij}$. 
A manifestly symmetric version of the energy-momentum tensor
can be derived from (\ref{emtensor}) and (\ref{maxmomergy}):
\begin{equation}\label{calt}
  {}^{\rm k}T_{\a\b}=-\,\lambda{}\,\hodge
  \left[{}\hodge(e_\a\rfloor F)\wedge(e_\b\rfloor F) 
  +{\frac{1}{2}}g_{\a\b} (\hodge F\wedge F)\right]\,.
\end{equation}

Thus ${}^{\rm k}T_{\a\b}$ is a {\it traceless symmetric} tensor(-valued 
0-form) with 9 independent components. Its symmetry is sometimes called 
a {\it bastard symmetry} since it interrelates two indices of totally
\index{electromagnetic!energy-momentum!symmetry} different origin,
as can be seen from (\ref{momergy4}). Without using a metric, the
symmetry cannot even be formulated, see (\ref{simax}).

\section{Concluding remarks}

The energy-momentum current 
\begin{equation}
  ^ {\rm k} \Sigma_\alpha :={\frac 1 2}\left[F\wedge(e_\alpha\rfloor
    H) - H\wedge (e_\alpha\rfloor F)\right]\,, 
\end{equation} 
emerging in the context of the fourth axiom, fulfills all the desirable
physical properties. If the Maxwell-Lorentz spacetime relation
$H=\sqrt{\varepsilon_0/\mu_0}\,\,^\star F$ is substituted, $\Sigmakin$
becomes symmetric and the conventional energy-momentum tensor for
vacuum electrodynamics is recovered. In a future paper, we will
discuss the consequences $\Sigmakin$ has for a consistently formulated
energy-momentum current {\em inside matter}. We hope to clarify
certain aspects of this age-old problem.

\bigskip

\nonumber {\bf Acknowledgments}. This work was supported by the 
DAAD (Kennziffer A/00/06508) and by the Alexander von Humboldt 
Foundation, Bonn. We are grateful for this support.

\end{document}